\documentclass[preprint]{aastex}

\newcommand{\refeqn}[1]{(\ref{#1})}
\newcommand{\reffig}[1]{Figure~\ref{#1}}
\newcommand{\reftbl}[1]{Table~\ref{#1}}
\newcommand{\refsec}[1]{Section~\ref{#1}}
\newcommand{\WMAP}{\textsl{WMAP}}

\newcommand{\aside}[1]{}
\newcommand{\muK}{{\mu {\rm K}}}
\newcommand{\VxV}{{V$\times$V}}
\newcommand{\VxW}{{V$\times$W}}
\newcommand{\WxW}{{W$\times$W}}

\shorttitle{\WMAP\ 5-year Angular Power Spectra}
\shortauthors{Nolta et al.}
\slugcomment{Accepted by ApJS}

\begin{document}
\title{Five-Year \textsl{Wilkinson Microwave Anisotropy Probe}
(\WMAP\altaffilmark{1}) Observations:\\ Angular Power Spectra}
\author{
{M. R. Nolta}   \altaffilmark{2},
{J. Dunkley} \altaffilmark{3,4,5},
{R. S. Hill}    \altaffilmark{6},
{G. Hinshaw} \altaffilmark{7},
{E. Komatsu} \altaffilmark{8},
{D. Larson}  \altaffilmark{9},
{L. Page}    \altaffilmark{4},
{D. N. Spergel} \altaffilmark{3,10},
{C. L. Bennett} \altaffilmark{9},
{B. Gold}    \altaffilmark{9},
{N. Jarosik} \altaffilmark{4},
{N. Odegard} \altaffilmark{6},
{J. L. Weiland} \altaffilmark{6},
{E. Wollack} \altaffilmark{7},
{M. Halpern} \altaffilmark{11},
{A. Kogut}   \altaffilmark{7}, 
{M. Limon}   \altaffilmark{12},
{S. S. Meyer}   \altaffilmark{13},
{G. S. Tucker}  \altaffilmark{14},
{E. L. Wright}  \altaffilmark{15}}
\altaffiltext{1}{WMAP is the result of a partnership between Princeton 
                 University and NASA's Goddard Space Flight Center. Scientific 
		 guidance is provided by the WMAP Science Team.}
\altaffiltext{2}{{Canadian Institute for Theoretical Astrophysics,  60 St. George St, University of Toronto,  Toronto, ON  Canada M5S 3H8}}
\altaffiltext{3}{{Dept. of Astrophysical Sciences,  Peyton Hall, Princeton University, Princeton, NJ 08544-1001}}
\altaffiltext{4}{{Dept. of Physics, Jadwin Hall,  Princeton University, Princeton, NJ 08544-0708}}
\altaffiltext{5}{{Astrophysics, University of Oxford,  Keble Road, Oxford, OX1 3RH, UK}}
\altaffiltext{6}{{Adnet Systems, Inc.,  7515 Mission Dr., Suite A100, Lanham, Maryland 20706}}
\altaffiltext{7}{{Code 665, NASA/Goddard Space Flight Center,  Greenbelt, MD 20771}}
\altaffiltext{8}{{Univ. of Texas, Austin, Dept. of Astronomy,  2511 Speedway, RLM 15.306, Austin, TX 78712}}
\altaffiltext{9}{{Dept. of Physics \& Astronomy,  The Johns Hopkins University, 3400 N. Charles St.,  Baltimore, MD  21218-2686}}
\altaffiltext{10}{{Princeton Center for Theoretical Physics,  Princeton University, Princeton, NJ 08544}}
\altaffiltext{11}{{Dept. of Physics and Astronomy, University of  British Columbia, Vancouver, BC  Canada V6T 1Z1}}
\altaffiltext{12}{{Columbia Astrophysics Laboratory,  550 W. 120th St., Mail Code 5247, New York, NY  10027-6902}}
\altaffiltext{13}{{Depts. of Astrophysics and Physics, KICP and EFI,  University of Chicago, Chicago, IL 60637}}
\altaffiltext{14}{{Dept. of Physics, Brown University, 182 Hope St., Providence, RI 02912-1843}}
\altaffiltext{15}{{UCLA Physics \& Astronomy, PO Box 951547,  Los Angeles, CA 90095-1547}}
\email{nolta@cita.utoronto.ca}

\begin{abstract}
We present the temperature and polarization angular power spectra of the
cosmic microwave background (CMB) derived from the first 5 years of
{\WMAP} data.
The 5-year temperature (TT) spectrum is cosmic variance limited up to multipole
$\ell=530$,
and individual $\ell$-modes have $S/N>1$ for $\ell<920$.
The best fitting six-parameter $\Lambda$CDM model
has a reduced $\chi^2$ for $\ell=33-1000$ of
$\chi^2/\nu=1.06$, with a probability to exceed of 9.3\%.
There is now significantly improved data near the third peak which 
leads to improved cosmological constraints.
The temperature-polarization correlation (TE) is seen with high significance.
After accounting for foreground emission, 
the low-$\ell$ reionization feature in the EE power spectrum 
is preferred by $\Delta\chi^2=19.6$ for optical depth
$\tau=0.089$ by the EE data alone,
and is now largely cosmic variance limited for $\ell=2-6$.
There is no evidence for cosmic signal in the BB, TB, or EB spectra after
accounting for foreground emission.
We find that, when averaged over $\ell=2-6$,
$\ell(\ell+1)C^{BB}_{\ell}/(2\pi) < 0.15\,\muK^2$ (95\% CL).
\end{abstract}


\section{Introduction}

%
%
%
%

The {\WMAP} satellite \citep{bennett/etal:2003} has measured the temperature and polarization
of the microwave sky at five frequencies from 23 to 94~GHz.
\cite{hinshaw/etal:prep}
presents our new, more sensitive temperature and polarization maps.
After removing a model of the foreground  emission from
these maps \citep{gold/etal:prep}, we obtain our
best estimates of the  temperature and polarization
angular power spectra of the cosmic microwave background (CMB).

This paper presents our statistical analysis of these CMB temperature and polarization maps.
Our basic analysis approach is similar to the approach described in our
first year {\WMAP} temperature analysis \citep{hinshaw/etal:2003} and polarization analysis \citep{kogut/etal:2003}
and in the three year {\WMAP} temperature \citep{hinshaw/etal:2007} and polarization analysis \citep{page/etal:2007}.
While most of the {\WMAP} analysis pipeline has been unchanged from our 3-year analysis, there
have been a number of improvements that have reduced the  systematic errors and increased the precision
of the derived power spectra.
\cite{hinshaw/etal:prep} describe the {\WMAP} data processing with an
emphasis on these changes.
\cite{hill/etal:prep} present our more complete analysis of the {\WMAP} beams
based on 5 years of Jupiter data and physical optics fits to both
the A- and B-side mirror distortions.
The increase in main beam solid angle leads to a revision in the beam function that impacts our computed power spectrum by raising the overall
amplitude for $\ell>200$ by roughly 2\%.
\cite{gold/etal:prep} introduce a  new set of masks that are designed to remove regions of free-free emission that were a minor (but detectable) contaminant
in analyses using the previous Kp2 mask used in the 1- and 3-year analysis.
\cite{wright/etal:prep} updates the
point source catalog presented in \cite{hinshaw/etal:2007} finding 67 additional sources.
\refsec{sec:changes} of this paper describes these changes and their implications for the measured power spectra.

\refsec{sec:tt} presents the temperature angular power spectrum (TT).
{\WMAP} has made a cosmic variance
limited measurement of the angular power spectrum to $\ell = 530$ and we now report results
into the ``third peak'' region.
The {\WMAP} results, combined with recent ground-based measurements
of the TT angular power spectrum
\citep{readhead/etal:2004,jones/etal:2006,reichardt/etal:prep}, result in
accurate measurements well into the ``fifth peak'' region.
For {\WMAP}, point sources are the largest astrophysical contaminant to the
temperature power spectrum.
We present estimates for the point source contamination based on
multi-frequency data, source counts and estimates from the bispectrum.

The polarization observations are decomposed into $E$ and $B$ mode components
\citep{kamionkowski/kosowsky/stebbins:1997,seljak/zaldarriaga:1997}.
Primordial scalar fluctuations generate only $E$ modes, while
tensor fluctuations generate both $E$ and $B$ modes.
With $T$, $E$ and $B$ maps, we compute the angular auto-power spectra of the
three fields, $TT$, $EE$ and $BB$,
and the angular cross-power spectra of these  three fields, $TE$, $TB$ and $EB$.
If the CMB fluctuations are Gaussian random fields, then these six angular
power spectra encode {\it all} of the statistical information in the CMB.
Unless there is a preferred sense of rotation in the universe,
symmetry implies that the $TB$ and $EB$ power spectrum are zero.
In \refsec{sec:tetb} we present both the TE and TB temperature-polarization
cross power spectra.
The {\WMAP} measurements of the TE spectrum now clearly see multiple peaks.
\aside{note from lyman: is this new? the second peak is a bit dicey}
The large angle TE anti-correlation is a distinctive signature of superhorizon
fluctuations \citep{zaldarriaga/spergel:1997}.
\cite{komatsu/etal:prep} discuss how the TB measurements constrain
parity-violating interactions.
\refsec{sec:eebb} presents both the EE and BB polarization power spectra.
The EE power spectrum now shows a clear $\sim 5 \sigma$ signature of cosmic
reionization.
\cite{dunkley/etal:prep} show that the amplitude of the signal implies that the
cosmic reionization was an extended process. \cite{dunkley/etal:prep} and
\cite{komatsu/etal:prep} discuss the cosmological implications of the angular
power spectrum measurements.

\section{Changes in the 5-year Analysis\label{sec:changes}}

The methodology used for the 5-year power spectra analysis is similar to
as that used for the 3-year analysis.
In this section we list the significant changes and their
impact on the results:

\begin{itemize}
\item \cite{hinshaw/etal:prep} describe the changes in
the map processing and the resultant reduction in the absolute calibration
uncertainty from 0.5\% to 0.2\%.

%

\item The temperature mask used to compute the power spectrum has been updated,
removing slightly more sky near the galactic plane, and more high-latitude
point sources \citep{gold/etal:prep}.
The galactic mask used in the 3-year and 1-year releases (Kp2) was constructed
by selecting all pixels whose K-band emission exceeded a certain threshold.
This procedure worked well in identifying areas contaminated by synchrotron
emission; however, it missed a few small regions contaminated by free-free,
particularly around $\rho$~Oph, the Gum nebula, and the Orion/Eridanus Bubble.
For the 5-year analysis we have constructed a new galactic mask to
remove these contaminated areas.

\cite{wright/etal:prep} updated the {\WMAP} point source catalog, finding
390 sources in the 5-year data, 67 more sources than in the 3-year catalog
\citep{hinshaw/etal:2007}.
Of these 67 new sources, 32 were previously unmasked, and therefore
added to the 3-year source mask to create the 5-year source
mask.\footnote{Six of the sources in the 5-year catalog were not added to
the mask; they were found in a late update to the catalog after the mask had
been finalized.}

All told, the new 5-year temperature power spectrum mask (KQ85)
retains 81.7\% of the sky,
while the 3-year mask (Kp2) retained 84.6\% of the sky.

\item
The 5-year polarization mask is the same as the 3-year P06 mask
described in \cite{page/etal:2007}, except that an additional 
0.27\% of the sky has been removed due to combining P06 with the new
processing mask \citep{hinshaw/etal:prep}.

\item
In addition to masking, the maps are further cleaned 
of galactic foreground emission using external templates.
The cleaning procedure is very similar to that of the 3-year analysis;
see \cite{gold/etal:prep} for details.
For the temperature map, three templates are used: 
a synchrotron template (the {\WMAP} $\rm K-Ka$ difference map),
an H$\alpha$ template as a proxy for free-free \citep{finkbeiner:2003},
and a thermal dust template \citep{finkbeiner/davis/schlegel:1999}.
For the polarization maps, two templates are used (since free-free is
unpolarized): the polarized K-band map and a polarized dust template
constructed from the unpolarized dust template, a simple model of the galactic
magnetic field, and polarization directions deduced from starlight.

\item
A great deal of work has gone into improving the determination of the 
beam maps and window functions \citep{hill/etal:prep}.
The main beam solid angles are larger than the 3-year estimates
by $\approx1$-2\% in V- and W-band.
Increased solid angle (i.e., greater map smoothing)
reduces the value of the transfer function $b_l$,
raising the deconvolved CMB power spectra.
The ratio of the 3-year to 5-year transfer functions can be seen in Figure~13 of
\cite{hill/etal:prep}; the net effect is to raise the TT power spectrum
by $\approx2\%$ for $\ell>200$, which is within the 3-year beam
$1\sigma$ confidence limits.
The beam transfer function uncertainty is smaller than the 3-year
uncertainty by a factor of $\approx2$.
The window function uncertainty is now
$\approx 0.6\%$ in $\Delta C_l/C_l$ for $200<\ell<1000$.

\end{itemize}

\section{Temperature Spectrum\label{sec:tt}}

The 5-year $\ell\le32$ spectrum is described in \cite{dunkley/etal:prep}.
At low-$\ell$ the likelihood function is no longer well approximated by a
Gaussian so that we explicitly sample the likelihood function to evaluate
the statistical distribution of each multipole.

We construct the 5-year TT spectrum for $\ell>32$ in the same fashion as
the 3-year spectrum; we refer the reader to \cite{hinshaw/etal:2007}
for details, and only briefly summarize the process as follows:

\begin{itemize}
\item We start with the single year V1,V2,W1--W4
resolution-10 maps,\footnote{12,582,912 pixels ($N_{\rm side} = 1024$).}
masked by the KQ85 mask, and further cleaned
via foreground template subtraction.
\item The pseudo-$C_l$ cross power spectra are computed for each pair
of maps.
Two weightings are used: flat weighting and inverse noise variance
($N_{\rm obs}$) weighting.
\item The year/DA cross power spectra are combined by band,
forming the \VxV, \VxW, and {\WxW} spectra.
The auto power spectra are not included in the combination, eliminating the
need to subtract a noise bias.
\item A model of the unresolved point source contamination
with amplitude $A_{ps}=0.011\pm0.001\,\muK^2{\rm sr}$
is subtracted from the band-combined spectra.
See \refsec{sec:ptsrc} for more details.
\item The \VxV, \VxW, \& {\WxW} spectra are optimally combined $\ell$-by-$\ell$
to create the final CMB spectrum.
\end{itemize}

As in the 3-year analysis, the diagonal elements of the $\hat C_l$ 
covariance matrix are calculated as
\begin{eqnarray}
(\Delta \hat C_l)^2 = \frac{2}{(2l+1)f^2_{\rm sky}(l)}(C_l+N_l)^2
\end{eqnarray}
where $C_l$ is the cosmic variance term and $N_l$ the noise term.
The value of $f_{\rm sky}(l)$, the effective sky fraction,
is calibrated from simulations:\footnote{
The Markov chains in \cite{dunkley/etal:prep} and \cite{komatsu/etal:prep}
were run with a version of the {\WMAP} likelihood code with older and 
slightly larger values for $f_{\rm sky}$.
The change in $f_{\rm sky}$ increased the TT errors by on average 2\%.
Rerunning the $\Lambda$CDM chain with the new $f_{\rm sky}$ leads to parameter
shifts of at most $0.1\sigma$.}
\begin{equation}
f_{\rm sky}(\ell) = \cases{
0.826 -0.091 (\ell/500)^2, & $\ell\le500$; \cr
0.777 -0.127 (500/\ell), & $\ell>500$.
}
\end{equation}

\begin{figure}
\plotone{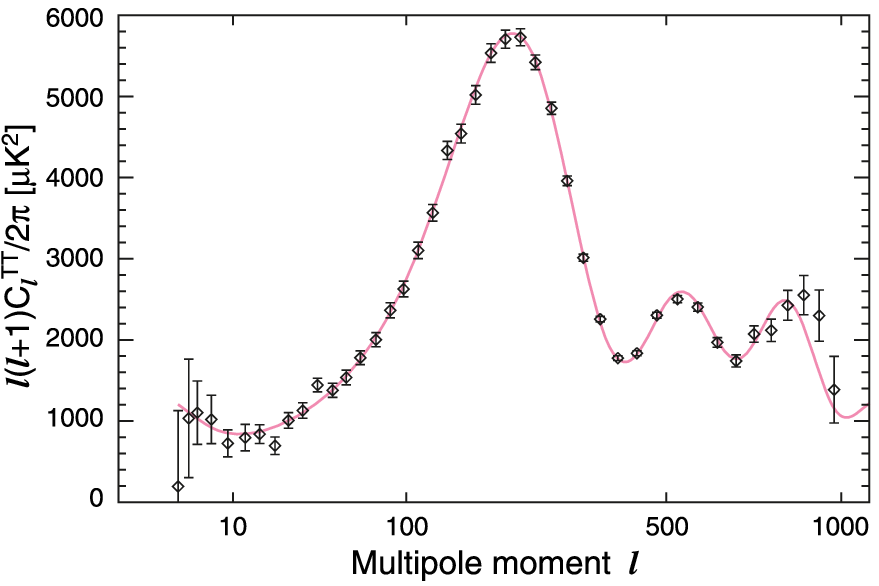}
\caption{\label{fig:tthighl}
The {\WMAP} 5-year temperature (TT) power spectrum.
The red curve is the best-fit theory spectrum from the $\Lambda$CDM/{\WMAP}
chain \citep[Table 2]{dunkley/etal:prep} based on {\WMAP} alone,
with parameters
$(
\ensuremath{\Omega_bh^2},
\ensuremath{\Omega_mh^2},
\ensuremath{\Delta_{\cal R}^2},
\ensuremath{n_s},
\ensuremath{\tau},
\ensuremath{H_0}
) = (
\ensuremath{0.0227},
\ensuremath{0.131},
\ensuremath{2.41},
\ensuremath{0.961},
\ensuremath{0.089},
\ensuremath{72.4}
)$.
The uncertainties include both cosmic variance, which dominates below
$\ell=540$, and instrumental noise which dominates at higher multipoles.
The uncertainties increase at large $\ell$ due to {\WMAP}'s
finite resolution. The improved resolution of the third peak near $\ell=800$
in combination with the simultaneous measurement of the rest of the spectrum
leads to the improved results reported in this release.
}
\end{figure}

The 5-year TT spectrum is shown in \reffig{fig:tthighl}.
With the greater S/N of the 5-year data the third acoustic peak is
beginning to appear in the spectrum.
The spectrum is cosmic variance limited up to $\ell=530$,
and individual $\ell$-modes have $S/N>1$ for $\ell<920$.
In a fit to the best cosmological $\Lambda$CDM model,
the reduced $\chi^2$ for $\ell=33-1000$ is
$\chi^2/\nu=1.06$, with a probability to exceed of 9.3\%.

\reffig{fig:ttunbinned} compares the unbinned 5-year TT spectrum
with the 3-year result. Aside from the small upward shift of the 5-year
spectrum relative to that of the 3-year, due to the new beam transfer function,
they are identical at low-$\ell$.
\reffig{fig:ttfreq} shows the unbinned TT
spectrum broken down into its frequency components (\VxV, \VxW, \WxW),
demonstrating that the signal is independent of frequency.

How much has the determination of the 3rd acoustic peak improved 
with the 5-year data?
Over the range $\ell=680-900$, which approximately spans the rise
and fall of the 3rd peak (from the bottom
of the 2nd trough to the point on the opposite side of the peak),
the fiducial spectrum is preferred over a flat mean spectrum
by $\Delta\chi^2=7.6$.
For the 3-year data it was $\Delta\chi^2=3.6$.
With a few more years of data, {\WMAP} should detect the curvature of
3rd peak  to greater than $3\sigma$.

\begin{figure}
\plotone{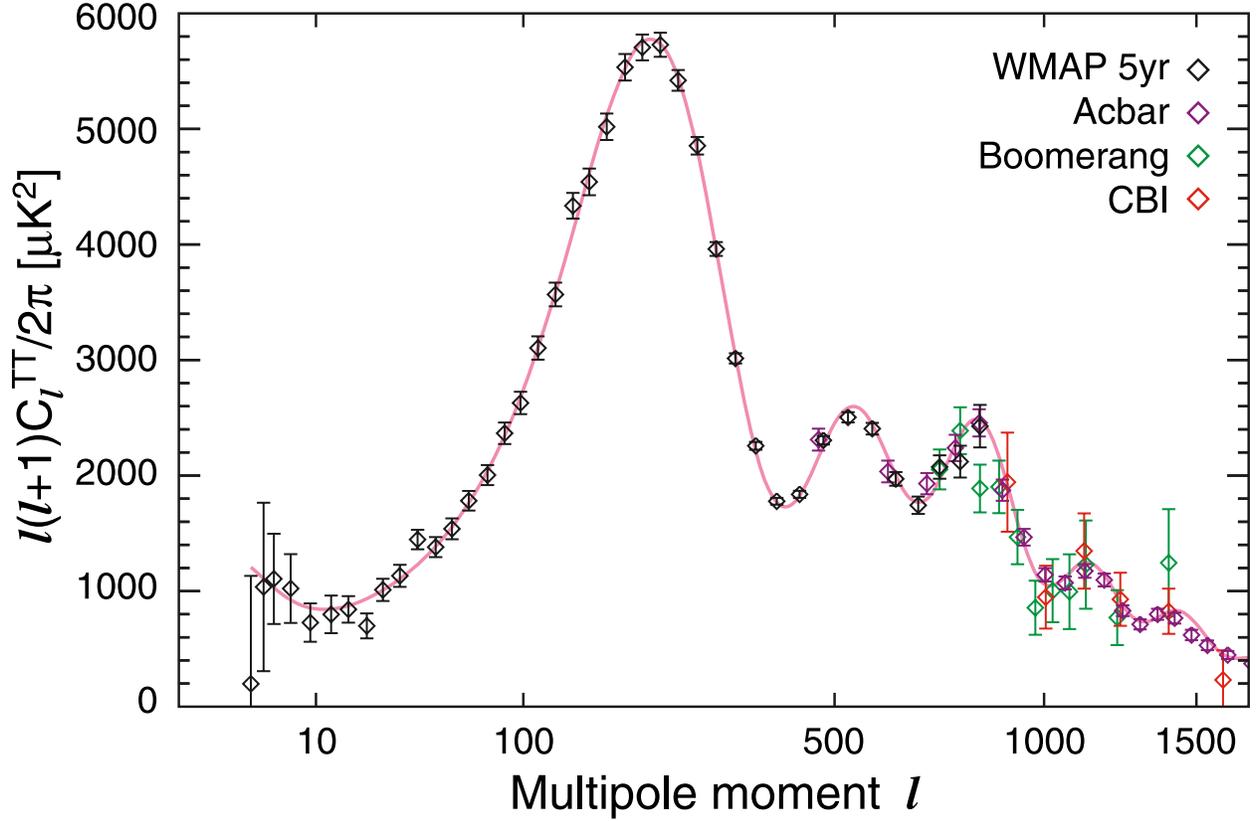}
\caption{\label{fig:tthighlvsothers}
The {\WMAP} 5-year TT power spectrum along with recent results from the
ACBAR \citep[purple]{reichardt/etal:prep},
Boomerang \citep[green]{jones/etal:2006},
and CBI \citep[red]{readhead/etal:2004} experiments. The other experiments
calibrate with {\WMAP} or {\WMAP}'s measurement of Jupiter (CBI).
The red curve is the best-fit $\Lambda$CDM model to the {\WMAP} data,
which agrees well with all data sets when extrapolated to higher-$\ell$.
}
\end{figure}

In \reffig{fig:tthighlvsothers} we compare the
{\WMAP} 5-year TT power spectrum along with recent results from other
experiments \citep{readhead/etal:2004,jones/etal:2006,reichardt/etal:prep},
showing great consistency between the various measurements.
Several on-going and future ground-based CMB experiments plan on calibrating
themselves off their overlap with {\WMAP} at the highest-$\ell$'s;
improving {\WMAP}'s determination of the 3rd peak will have the added benefit
of improving their calibrations.

\subsection{Unresolved Point Source Correction\label{sec:ptsrc}}

A population of point sources, Poisson-distributed over the sky, contributes
an additional source of white noise to the measured TT power spectrum,
$C^{TT}_l \to C^{TT}_l + C^{ps}$.
Given a known source distribution $N(>S)$, the number of sources per steradian
with flux greater than $S$, the point-source induced signal is
\begin{eqnarray}
C^{ps} = g(\nu)^2 \int^{S_c}_0{dS\,\frac{dN}{dS}S^2} \qquad[\muK^2{\rm sr}]
\label{eqn:Cps}
\end{eqnarray}
where $S$ is the source flux, $S_c$ is the flux cutoff (above which sources
are masked and removed from the map), and $g(\nu)=(c^2/2k\nu^2)r(\nu)$
converts flux density to thermodynamic temperature, with
\begin{eqnarray}
r(\nu)=\frac{(e^x-1)^2}{x^2e^x},
\quad x\equiv h\nu/kT_{\rm CMB}
\end{eqnarray}
converting antenna to thermodynamic temperature.

At the frequencies and flux densities relevant for {\WMAP}, source counts
are dominated by flat-spectrum radio sources, which have flux spectra
that are nearly constant with frequency
($S\sim\nu^\alpha$ with $\alpha\approx0$).
\cite{wright/etal:prep} finds the average spectral index of sources
bright enough to be detected in the {\WMAP} 5-year data to be
$\langle\alpha\rangle=-0.09$, with an intrinsic 
dispersion of $\sigma_\alpha = 0.176$.
Since a source with flux $S\sim\nu^\alpha$ has a thermodynamic temperature
$T\sim\nu^{\alpha-2}r(\nu)$,
we model the frequency dependence of $C^{ps}$ as
\begin{eqnarray}
C^{ps}({\nu_i,\nu_j}) = A_{ps} r(\nu_i)r(\nu_j)
\left(\frac{\nu_i\nu_j}{\nu^2_Q}\right)^{\alpha-2}
\label{eqn:srcmodel}
\end{eqnarray}
where $\nu_{i,j}$ are the frequencies of the two maps used to calculate
the TT spectrum, $A_{ps}$ is an unknown amplitude,
and $\nu_Q=40.7\,{\rm GHz}$ is the Q-band central frequency.

In this section, we estimate the value of $A_{ps}$ needed to correct the TT
power spectrum, finding $A_{ps} = 0.011\pm0.001\,\muK^2{\rm sr}$,
and discuss incorporating its uncertainty into the likelihood function.

\subsubsection{Estimating the correction}

For a fixed beam size, flat-spectrum radio sources are much fainter in
the W-band temperature maps than in Q- or V-band,
allowing us to use the frequency dependence of the TT spectrum at high-$\ell$
to constrain the value of $A_{ps}$.
As in previous releases, the estimator we use is
\begin{eqnarray}
\hat A_{ps} &=& \frac{\sum_{l\alpha\beta} C^{\alpha}_l(\Sigma^{-1})^{\alpha\beta}_lh^\beta_l}{\sum_{l\alpha\beta} s^{\alpha}_l(\Sigma^{-1})^{\alpha\beta}_lh^{\beta}_l}
\label{eqn:Apsestimator} \\
h^{\gamma}_l & = & s^{\gamma}_l - \frac{\sum_{\alpha\beta}s^{\alpha}_l(\Sigma^{-1})^{\alpha\beta}_l}{\sum_{\alpha\beta}(\Sigma^{-1})^{\alpha\beta}_l}
\end{eqnarray}
where greek letters represent a pair of frequencies (e.g., VW),
$C^{\alpha}_l$ is the measured TT cross-power spectrum,
$\Sigma^{\alpha\beta}_l$ is the $\langle C^{\alpha}_lC^{\beta}_l\rangle$
covariance matrix including cosmic variance and detector noise,
and $s^{\alpha}_l = l(l+1)C^{ps}(\alpha)/2\pi$.
The inverse estimator variance ($[\delta\hat A_{ps}]^{-2}$)
is given by the denominator of \refeqn{eqn:Apsestimator}.
While $\Sigma^{\alpha\beta}_l$ does not include the off-diagonal
coupling due to the mask, the diagonal elements are renormalized to
account for the loss of sky coverage.

\begin{figure}
\plotone{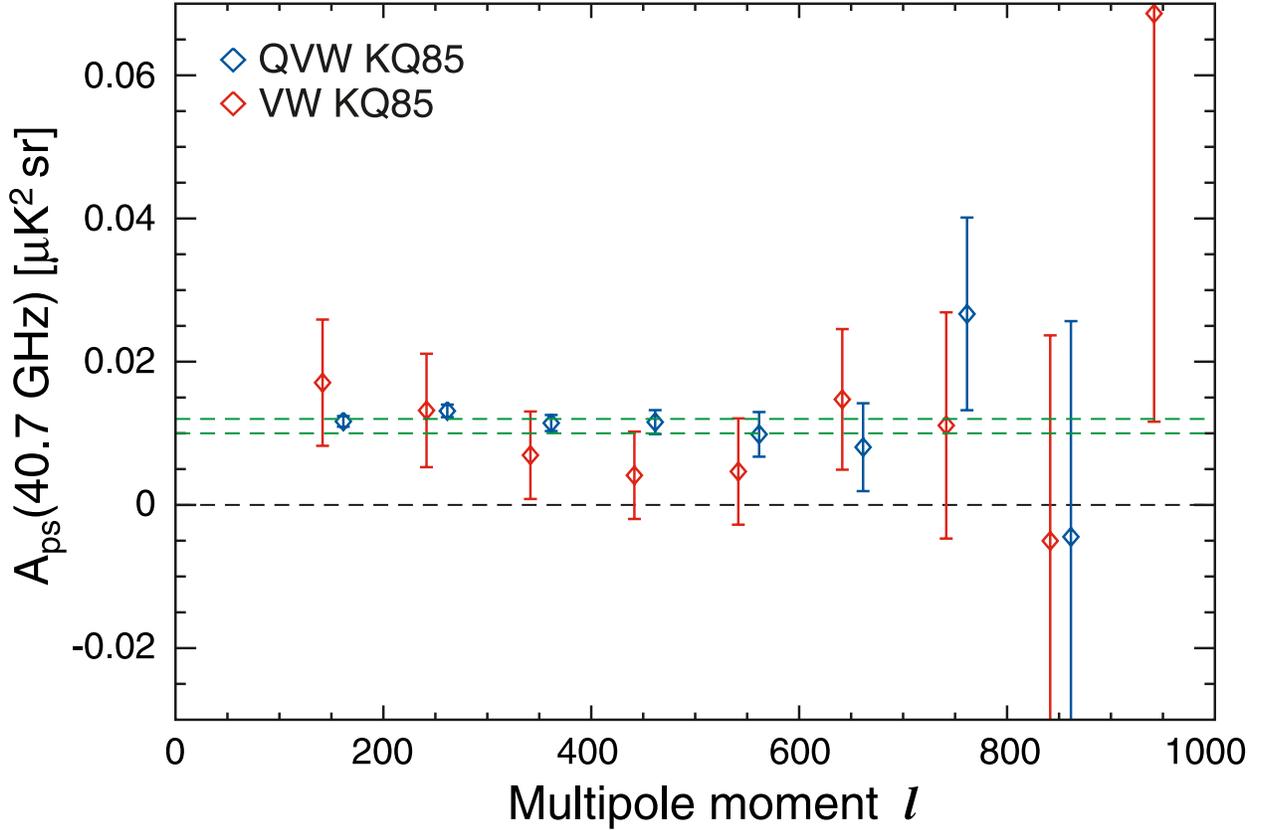}
\caption{\label{fig:ptsrc}
The unresolved point source contamination $A_{ps}$,
measured in bins of $\Delta\ell=100$ evaluated at 40.7 GHz (Q-band).
For a source population whose fluxes are independent of frequency
$A_{ps}$ scales roughly as $\sim\nu^{-2}$ in the {\WMAP} data.
The red data points are from the analysis of V and W bands
alone and the blue points are from the analysis of Q, V, and W bands.
The horizontal dashed green lines, at 0.010 and 0.012,
show the $1\sigma$ bounds for our adopted value of $A_{ps}$.
Note that the QVW amplitude is independent of $\ell$.
}
\end{figure}

\begin{figure}
\plotone{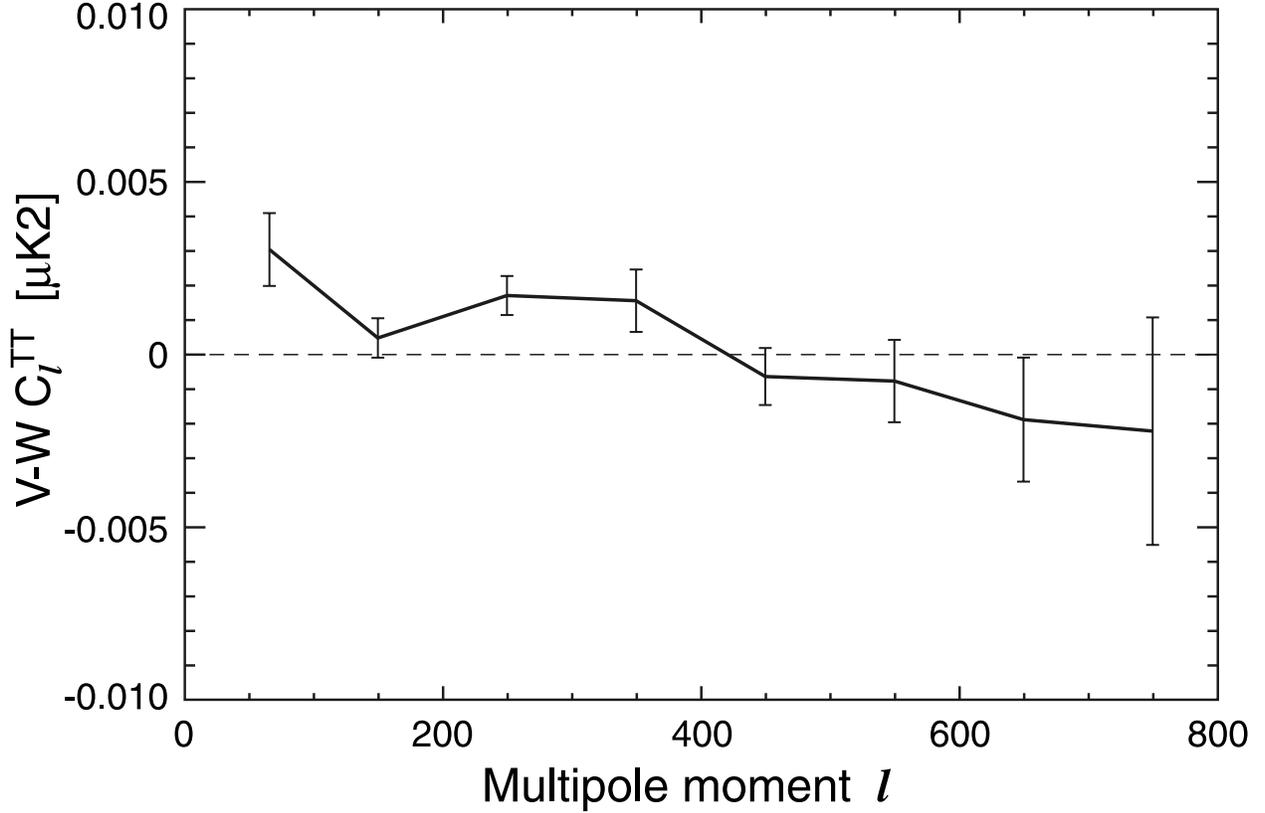}
\caption{\label{fig:ttnull}
The TT $\rm V-W$ null spectrum. After correcting for unresolved point
source emission, the individual power spectra are subtracted
in power spectrum space. The result is consistent
with zero and thus there is no evidence of point source contamination.
In these units, point source contamination would be evident as a horizontal
offset from zero.
At $\ell=500$, the TT power spectrum is $C^{\rm TT}_\ell\approx0.06$;
thus the contamination is limited to roughly 3\% in power.
}
\end{figure}

Measured values for $A_{ps}$ are listed in \reftbl{tbl:Aps} for various
frequency combinations (QVW \& VW) and galactic masks (KQ85, KQ80, \& KQ75).
The QVW estimates are insensitive to the galactic mask; the VW estimate
increases somewhat as more of the sky is masked.
Both the QVW and VW estimates prefer the same value
($\approx0.011\,\muK^2{\rm sr}$)
of $A_{ps}$ when the KQ75 mask is used.
While we restrict the data to $\ell=300-800$,
the QVW estimate is only a weak function of the chosen $\ell$-range;
\reffig{fig:ptsrc} shows $A_{ps}$ estimated in bins of width $\Delta\ell=100$.
We adopt $A_{ps}=0.011\pm0.001\,\muK^2{\rm sr}$ as our correction to the
final combined TT spectrum.
The consistency between $\ell$-bins and between QVW and VW seen
in \reffig{fig:ptsrc} is an important null test for the angular power spectrum.
$A_{ps}$ (VW) is proportional to the power in the (V-W) map in a given
$\ell$-range.
\reffig{fig:ttnull}
shows no evidence for any detectable residual signal in the VW maps after point source subtraction.

Because radio sources can only have positive flux they introduce a positive
skewness to the maps, which can be detected in searches for non-Gaussianity.
\cite{komatsu/etal:prep} estimated the bispectrum induced by sources,
finding $b^{ps} = (4.3\pm1.3)\times10^{-5}\,\muK^3{\rm sr}^2$ at Q-band.
\aside{note from David: I am nervous that our point source selection at
K band slightly skews the maps toward a negative value that leads to an
underestimate of $b^{ps}$}
Is this consistent with the value of $A_{ps}$ measured from the power
spectrum?
Given a theoretical model for the source number counts $N(>S)$, one can
predict the measured values of $C^{ps}$ and $b^{ps}$.
Several models exist in the literature; we tested our results against two,
\citet[Tof98]{toffolatti/etal:1998}\footnote{In \cite{bennett/etal:2003}
we found that the Tof98 model needed to
be rescaled by a factor of 0.66 to match the {\WMAP} 1-year number counts; 
\cite{wright/etal:prep} refined the rescaling factor to 0.64 to match the
{\WMAP} 5-year source counts.}
and \citet[deZ05]{dezotti/etal:2005}.
$C^{ps}$ is calculated via \refeqn{eqn:Cps}, and $b^{ps}$ from
\begin{eqnarray}
b_{\rm src} = g^3(\nu) \int_0^{S_c}{dS\,\frac{dN}{dS}S^3} .
\end{eqnarray}
where $g(\nu)$ and $S_c$ are defined in \refeqn{eqn:Cps}.
The comparison is complicated by the fact that $S_c$ is unknown.
We mask out not only the sources detected in {\WMAP} data, but 
also undetected sources from external catalogues that are likely to
contribute contaminating flux.
However, a single value of $S_c$ predicts both $C^{ps}$ and $b^{ps}$, 
so we can in principle tune $S_c$ to match one, and see if it agrees with the
other.
In \reftbl{tbl:ptsrc} we compare our measured values of $C^{ps}$
and $b^{ps}$ with the rescaled Tof98 and deZ05 predictions for 
several values of $S_c$.
There is some tension between the measured values and the model predictions.
Given our measured value for $b_{ps}$ the models would prefer a smaller
value for $A_{ps}$, in the range 0.008-$0.010\,\muK^2{\rm sr}$.
For the Tof98 model, the $S_c\approx0.52$ predictions
are within $1\sigma$ of both $C^{ps}$ and $b^{ps}$.
However, the deZ05 model appears to be discrepant, and a single value for $S_c$
cannot match both $C^{ps}$ and $b^{ps}$.

Other groups have independently estimated the unresolved source contamination,
and their results are in general agreement with ours.
When the 3-year data was initially released the correction was
$A_{ps}=0.017\pm0.002\,\muK^2{\rm sr}$.
\cite{huffenberger/eriksen/hansen:2006}
reanalyzed the data and claimed $A_{ps}=0.011\pm0.001$,
noticing that $A_{ps}$ was sensitive to the choice of galaxy mask;
using the Kp0 mask instead of Kp2 reduced the value of $A_{ps}$.
Revisiting our original estimate for the 3-year analysis,
we reduced the correction to $0.014\pm0.003$ for the published papers.
In a subsequent paper, \cite{huffenberger/etal:2007prep}, the same group
corrected their original estimate after finding a small error, finding
$0.013\pm0.001$, consistent with our published result.

\section{Temperature-Polarization Spectra\label{sec:tetb}}


%



The standard model of adiabatic primordial density fluctuations 
predicts a correlation between the temperature and polarization 
fluctuations. The temperature traces 
primarily the density, and E-mode polarization the velocity, 
of the photon-baryon plasma at recombination. 
The correlation was seen in earlier WMAP data by 
\cite{kogut/etal:2003} and \cite{page/etal:2007}. The anti-correlation 
near $\ell=30$ provides evidence that fluctuations exist on superhorizon
scales, as it is observed on an angular scale larger than the acoustic 
horizon at decoupling \cite{spergel/zaldarriaga:1997}.

No significant changes have been made in the five-year TE analysis. 
We continue to use the method described in \cite{page/etal:2007} to 
compute the TE power spectrum. The inputs are the 
KaQV polarization maps \citep{gold/etal:prep}, and the 
VW temperature maps. For high multipoles 
$\ell>23$, the likelihood can be approximated 
as a Gaussian, and we continue to use the ansatz given in 
Appendix C of \cite{page/etal:2007} to compute the covariance matrix.
At low multipoles, $\ell \le 23$, the likelihood of the 
polarization data is evaluated directly from the maps, 
following Appendix D in \cite{page/etal:2007}.

\begin{figure}
\plotone{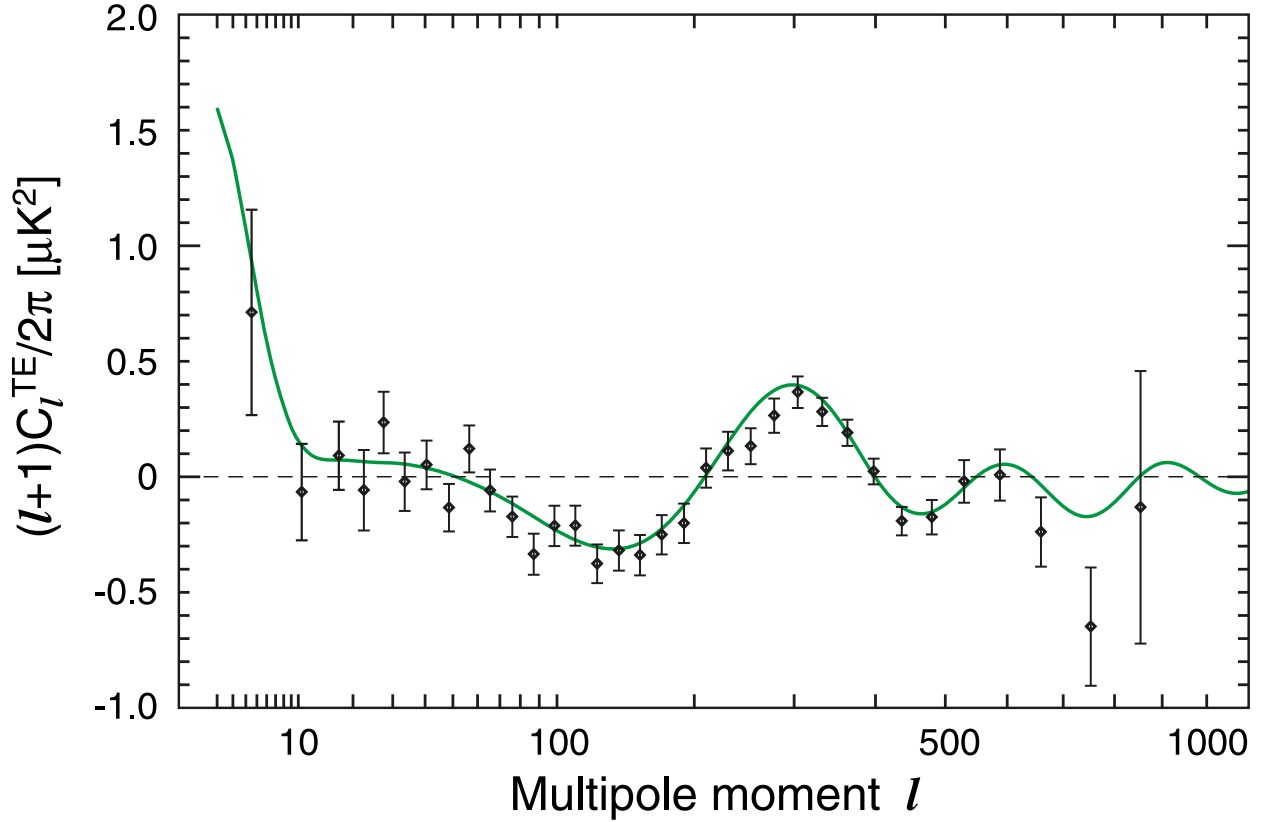}
\caption{\label{fig:tehighl}
The {\WMAP} 5-year TE power spectrum.
The green curve is the best-fit theory spectrum from the $\Lambda$CDM/{\WMAP}
Markov chain \citep{dunkley/etal:prep}.
For the TE component of the fit, $\chi^2=415$, and there are
427 multipoles and 6 parameters; thus the number of degrees of freedom is
$\nu=421$,  leading to $\chi^2/\nu=0.99$.
The particle horizon size at decoupling corresponds to $l\approx100$.
The clear anticorrelation between the primordial plasma density
(corresponding approximately to T) and velocity (corresponding approximately
to E) in causally disconnected regions of the sky indicates that the
primordial perturbations must have been on a superhorizon scale.
Note that the vertical axis is $(\ell+1)C_\ell/(2\pi)$, and not
$\ell(\ell+1)C_\ell/(2\pi)$.
}
\end{figure}

\reffig{fig:tehighl} shows the TE spectrum.  At low-$\ell$ the 
spectrum and error bars are approximated using the Gaussian form, although
these are not used for cosmological analysis. With five years of data the 
anti-correlation at $\ell=140$ is clearly seen in the data, and the 
correlation at $\ell=300$ is measured with higher accuracy. 
The second anti-correlation at $\ell \sim 450$ is now better characterized, 
and is consistent with predictions of the $\Lambda$CDM model.
The structure tests the consistency of the simple model, which fits 
both the TT and TE spectra with only six parameters. The best-fit
$\Lambda$CDM model has $\chi^2=415$ for the TE component, with 
421 degrees of freedom, giving $\chi^2/\nu=0.99$. 
The consistency confirms that the 
fluctuations are predominantly adiabatic, and 
constrains the amplitude of isocurvature modes.

The signal at the lowest multipoles, evaluated using the 
exact likelihood, is used to provide additional 
constraints on the reionization history. Although small, the 
measurement is consistent with the EE signal, and consistent with 
the three-year WMAP observations \cite{page/etal:2007}. 
\cite{dunkley/etal:prep} discuss constraints on reionization.

\begin{figure}
\plotone{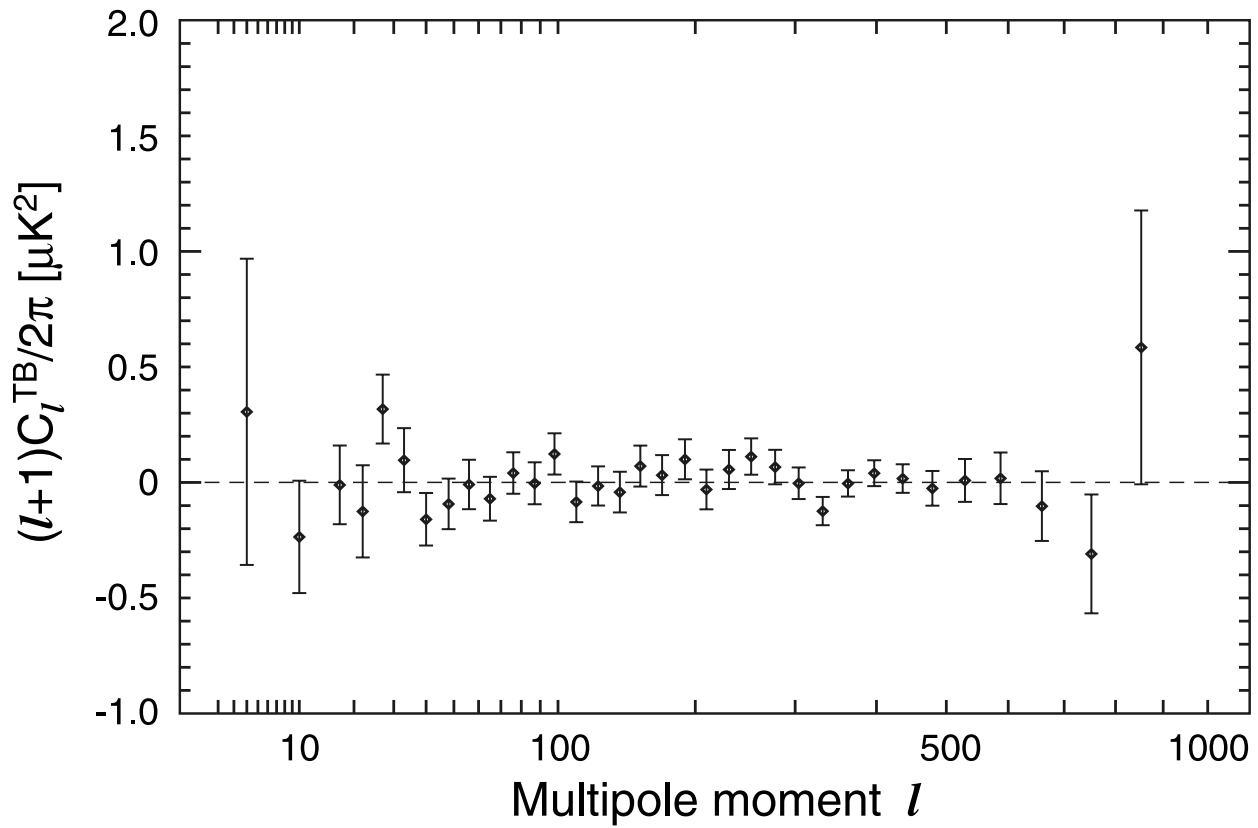} 
\caption{\label{fig:tbhighl}
The {\WMAP} 5-year TB power spectrum, showing no evidence of cosmological
signal.
The null reduced $\chi^2$ for $\ell=24-450$ is 0.97.
Note that the vertical axis is $(\ell+1)C_\ell/(2\pi)$, and not
$\ell(\ell+1)C_\ell/(2\pi)$.
}
\end{figure}

No correlation is expected between the temperature and the 
B-mode polarization. The TB spectrum is 
therefore primarily used as a null test, and is shown in 
\reffig{fig:tbhighl}. It is consistent with no signal, as expected;
over $\ell=24-450$ the reduced null $\chi^2$ is 0.97.
This measurement is used in \cite{komatsu/etal:prep} 
to place constraints 
on the presence of any parity violating terms coupled to 
photons, that could produce a TB correlation. 
We now include the TB spectrum at high $\ell$ as an optional module for the 
likelihood code.

\section{Polarization Spectra\label{sec:eebb}}

Due to its thermal stability \citep{jarosik/etal:2007}
and well-characterized gain,
{\WMAP} can measure polarization signals even though the scan pattern was not
optimized for doing so.
%
The polarization signal is manifested in the time ordered data (TOD) differently
from the temperature signal.
As a result, some of the low-$\ell$ polarization multipoles
are well sampled and other multipoles are poorly sampled and have large
statistical errors \citep{hinshaw/etal:2007,page/etal:2007}.
This is a rather different situation than from that of the temperature
spectrum, and the data must be analyzed with some care.



When we analyze the $\ell=2$ temperature power spectrum, we use the likelihood
function rather than Gaussian errors, as the Gaussian approximation starts to
break down with only $\approx4$ effective modes measured in the map
(the reduction is due to $f_{\rm sky}\approx0.7$).
For polarization, this effect is even more dramatic, as our scan pattern
significantly lowers the effective number of multipoles measured,
particularly for EE $\ell=2$, 5, 7 and 9 and BB $\ell=3$ 
\citep[the peaks seen in Figure 16 in][]{page/etal:2007}. 
\reffig{fig:simplebb} demonstrates the importance of using the full likelihood
description.
The Figure shows both the pseudo-$C_l$ estimates of the
$\ell=2-7$ BB multipoles and the conditional likelihoods
computed using the {\WMAP} likelihood code by varying the multipole in 
question, keeping the rest of the spectrum fixed to the
fiducial best-fit $\Lambda$CDM model.
From the plots it is clear that the best estimates of
the mean and the uncertainty are not attained with the pseudo-$C_l$ estimates.

We next consider the low-$\ell$ EE and BB power spectra in more detail.
The low-$\ell$ EE power spectrum is shown in \reffig{fig:eelowl}.
The uncertainties are obtained from the  conditional likelihood
and include cosmic variance;
thus  one  cannot  double the error flags to  get  the  95\% confidence limits.
If we zero out the $\ell<10$ portion of the fiducial EE \& TE spectra
the $\chi^2$ increases by 22.3, of which 2.7 is due to TE.
Thus the reionization feature in the EE power spectrum
is preferred by $\Delta\chi^2=19.6$.
The $\ell=2$, 3, 4, \& 6 multipoles are cosmic variance limited, and
the S/N ratio for the combined $\ell=2-7$ bandpower is 11.

Considerable effort has gone into understanding the W-band $\ell=7$ EE signal.
Because of the apparent anomalously high $\ell=7$ EE value computed by the
pseudo-$C_\ell$ algorithm, we have avoided using the W-band maps in
cosmological analysis and use them only as an additional check on
various models. Figure~8 of \cite{hinshaw/etal:prep} shows that the $\ell=7$
value, while high, appears to be consistent with being in the tail of a properly
computed likelihood distribution.
The W-band $\ell=7$ problem may be a signature of
poor statistics rather than a systematic.
However, more data are needed to understand this potential anomaly.
The $\ell=3$ BB signal gives perhaps the clearest example of the importance
of using the full likelihood code.
While the pseudo-$C_l$ estimate implies a significant
detection of power, the full likelihood code shows this to not be the case.
The physical cause of the large uncertainty is that with our scan strategy
an $\ell=3$ BB signal resembles an offset in the data and thus is not well
separated from the baseline \citep{page/etal:2007,hinshaw/etal:prep}.

We see no evidence for a B-mode signal at low $\ell$, limiting the possible
level to $\ell(\ell+1)C^{BB}_{\ell=2-6}/(2\pi) < 0.15~\mu$K$^2$ (95\% CL),
including cosmic variance. With $\tau= 0.1$ and $r=0.2$, a typical estimate for currently favored models of inflation, $\ell(\ell+1)C^{BB}_{\ell=2-6}/(2\pi)
\approx0.008~\mu$K$^2$.
Since a signal of $0.15~\mu$K$^2$ corresponds roughly to $r\approx20$,
one can see that {\WMAP}'s limit is not based on the BB data, but on the tensor
contribution to the
TT and EE spectra as discussed in \cite{komatsu/etal:prep}.

For EE at $\ell>10$, there are hints of signal in the data consistent with the
standard $\Lambda$CDM model. However, the significance is not great enough to
contribute to knowledge of the cosmological parameters.
The 5-year high-$\ell$ EE spectrum is shown in \reffig{fig:eehighl},
along with recent results from ground-based experiments
\citep{leitch/etal:2005,montroy/etal:2006,sievers/etal:2007}.
For $\ell=50-800$, $\chi^2=859.1$ assuming $C^{EE}_l=0$,
and drops by 8.4, or almost $3\sigma$, assuming the
standard $\Lambda$CDM model.
For the 3-year data the equivalent change in $\chi^2$ was 6.2.

The high-$\ell$ BB spectrum is consistent with no signal,
having a reduced $\chi^2$ of 1.02 over $\ell=50-800$ for the QV data.
The lack of any signal in the low and high $\ell$ BB data is a necessary
check of the foreground subtraction. As seen in Page et al. (2007),  foreground
emission produces E-modes and B-modes at similar levels; thus the absence of a
B-mode signal suggests that the level of contamination in the  E-mode
signal is low.
This is quantified in \cite{dunkley/etal:prep}.

\section{Summary and Conclusions\label{sec:summary}}

We have presented the temperature and polarization angular power spectra of the
cosmic microwave background (CMB) derived from the first 5 years of
{\WMAP} data.
With greater integration time our determination of the third acoustic
peak in the TT spectrum has improved.
The low-$\ell$ reionization feature in the EE spectrum is now detected
at nearly $5\sigma$.
The TB, EB, \& BB spectra show no evidence for cosmological signal.
The spectra are in excellent agreement with the best
fit $\Lambda$CDM model.
Our knowledge of the power spectrum is improving
both due to more detailed analyses, better modeling and understanding of the
foreground emission, and more integration time.

All of the 5-year {\WMAP} data products are being made available through the
Legacy Archive for Microwave Background Data Analysis
(LAMBDA\footnote{\url{http://lambda.gsfc.nasa.gov/}}),
NASA's CMB Thematic Data Center.
The temperature and polarization angular power spectra presented here are
available, as is the WMAP likelihood code which
 incorporates our estimates of the Fisher matrix, point sources and
beam uncertainties. 

\acknowledgements

The {\WMAP} mission is made possible by the support of the Science Mission
Directorate Office at NASA Headquarters.
This research was additionally supported by NASA grants NNG05GE76G,
NNX07AL75G S01, LTSA03-000-0090, ATPNNG04GK55G, and ADP03-0000-092.
EK acknowledges support from an Alfred P. Sloan Research Fellowship.
This research has made use of NASA's Astrophysics Data System Bibliographic
Services.
We acknowledge use of the CAMB, CMBFAST, CosmoMC, and HEALPix
\citep{gorski/etal:2005} software packages.

\appendix

\section{Likelihood treatment of source/beam uncertainties\label{sec:appsrclike}}

In this section, we test the treatment of the unresolved source
correction and beam uncertainties in the {\WMAP} likelihood code,
and show that it produces the correct results for cosmological 
parameters.

We adopt the same likelihood treatment of the unresolved point source
correction uncertainty for the 5-year likelihood code as used in the 3-year code
\citep[Appendix A]{hinshaw/etal:2007}, updated for the 5-year value of $A_{ps}$.
Briefly, a correction to the logarithmic likelihood,
${\cal L}\equiv -2\ln L= {\cal L}_0 + {\cal L}_1$
where ${\cal L}_0$ is the standard likelihood and ${\cal L}_1$ the 
combined source \& beam correction,
is calculated assuming the $C_l$ are normally distributed, a reasonable
assumption at high $\ell$.

\citet[Huf08]{huffenberger/etal:2007prep} disagreed with
the source \& beam likelihood module used in the 3-year analysis,
pointing out that the uncertainty in $n_s$ (the index of primordial scalar
perturbations) was unchanged even if the uncertainty in $A_{ps}$ was
increased by a factor of 100 (Fig. 2 in their paper).
They proposed an alternative approach,
integrating the beam/point source covariance
matrix into the cosmic variance/noise/mask covariance matrix and inverting the
result in order to compute ${\cal L}$ directly,
instead of calculating ${\cal L}_1$ as a separate correction.
Using this form of the likelihood, as $\delta A_{ps}$ was increased,
the uncertainty in $n_s$ increases (albeit modestly; $\delta n_s$ increased
by 38\% when $\delta A_{ps}\to100\times \delta A_{ps}$).

However, while we agree that it is striking that the error in $n_s$
is seemingly unaffected by the uncertainty in $A_{ps}$, we have some concerns
regarding the Huf08 approach.
To quote Huf08, ``the errors on the source measurement do not make much
difference, as long as $[\delta A_{ps}] < 0.003\,[\muK^2{\rm sr}]$'', and their Fig.~2 implies
the same holds true when $\delta A_{ps}=0.003$. This value is significant,
because it is the uncertainty adopted for the 3-year {\WMAP} analysis.
When Huf08 adopted the same uncertainty,
they found the same absolute uncertainty in 
$n_s$ as the {\WMAP} team, but their central value was shifted higher by 0.005.
This shift persisted as $\delta A_{ps}\to0$, and thus was seemingly not due to
the point source uncertainty. The conclusion we draw is that they
found the same value of $\delta n_s$ as the {\WMAP} 3-year analysis,
but their value of $n_s$ was biased high because of the way they treat the
beam uncertainties.
We believe the Huf08 value of $n_s$ would be in agreement with that found in
WMAP3, but that it is biased high due to their treatment of beam uncertainties.

Huf08 quoted the value of ${\cal L}_1$ computed with their
alternative likelihood module for a particular CMB spectrum distributed with
the {\WMAP} 3-year likelihood code test program,
finding ${\cal L}_1=-2.64$, whereas the {\WMAP} value is ${\cal L}_1=-1.22$.
As a check, we numerically marginalize the ${\cal L}_0$ portion of the
likelihood over beam and point
source errors, to see if we can reproduce their value.
The desired integral is
\begin{eqnarray}
\exp(-{\cal L}_1/2)
&=& \frac{1}{L_0(d|C_l)}\int{dxd\vec{y}\, 
e^{-(x^2 + \vec{y}^T\vec{y})/2}
L_0(d|C_l(x,\vec{y})) }
\end{eqnarray}
where
\begin{eqnarray}
C_l(x,\vec{y}) \equiv \left(C^{TT}_l + x \sigma^{\rm ptsrc}_l\right)
\left(1 + \sum_{i}y_i \sigma^{\rm beam}_l(i)\right),
\end{eqnarray}
is the theoretical model ($C^{TT}_l$) perturbed by point source and
beam errors.
With 10 dimensions to integrate over (nine beam modes and one point source
mode), normal grid-based quadrature is impractical,
so we turn to Monte Carlo integration instead:
\begin{eqnarray}
\exp(-{\cal L}^{MC}_l/2) &\approx& \frac{1}{N_{MC}}\sum_{i=1}^{N_{MC}}{e^{\ln L(d|C_l(x^{(i)},\vec{y}^{(i)}))-\ln L(d|C_l)}}
\end{eqnarray}
where $x^{(i)}$ and $y^{(i)}_j$ are independent unit-variance normal deviates.
With $N_{MC}=10^4$ points, we find ${\cal L}^{MC}_1=-1.29\pm0.04$,
consistent with the {\WMAP} result of $-1.22$, but not the
Huf08 result of $-2.64$.

As a further test of whether the our cosmological parameter
estimates fully capture the point source uncertainty,
we have run a Markov chain with a modified form of the point source likelihood
module, dubbed SRCMARG.
The point source correction is calculated via a simple numerical integration,
\begin{eqnarray}
\exp(-{\cal L}^{\rm ptsrc}_1/2) &=&
\int{d\alpha\,\frac{1}{\sqrt{2\pi}}e^{-\alpha^2/2} L_0(d|C_l+\alpha\sigma^{\rm ptsrc}_l)}
\label{eqn:srcmarg}\\ &\approx&
\frac{\Delta}{\sqrt{2\pi}}
\sum^{N}_{i=-N}w_i e^{-(i\Delta)^2/2} L_0(d|C_l+i\Delta\sigma^{\rm ptsrc}_l)
\label{eqn:srcquadrature}
\end{eqnarray}
with $N=25$, $\Delta=0.2$, and $w_i=1$ for except at the endpoints
where $w_{|N|}=1/2$ (the trapezoidal rule).
The resulting one-dimensional marginalized distribution for $\sigma_8$,
shown in the left panel of \reffig{fig:nssig8}, is
indistinguishable from our standard result.
We have also run a SRCMARG chain with the error
increased by a factor of $5$ (i.e., $\delta A_{ps}=0.005$).
In this case the uncertainty in $\sigma_8$ increases by 15\%.

Likewise, we have run similar tests of the beam uncertainty, dubbed BEAMMARG.
The approach is the same as SRCMARG, but with
``$C_l+\alpha\sigma^{\rm ptsrc}_l$'' in \refeqn{eqn:srcmarg}
replaced by ``$C_l(1+\alpha\sigma^{\rm beam}_l)$'',
where $\sigma^{\rm beam}_l$ is the noisiest beam eigenmode, shown in
Figure 12 of \cite{hill/etal:prep}.
The 1D marginalized distributions for $n_s$ are shown in the right panel of
\reffig{fig:nssig8}.
As with SRCMARG, the BEAMMARG result is indistinguishable from our
standard result.
Inflating the beam error by a factor of 20 results in a 14\% increase
in $\delta n_s$, along with a slight shift in $n_s$ away from unity.



\begin{thebibliography}{31}
\expandafter\ifx\csname natexlab\endcsname\relax\def\natexlab#1{#1}\fi

\bibitem[{{Bennett} et~al.(2003)}]{bennett/etal:2003}
{Bennett}, C.~L., et~al. 2003, \apj, 583, 1

\bibitem[{{Bischoff} et~al.(2008)}]{bischoff/etal:2008}
{Bischoff}, C., et~al. 2008, ArXiv e-prints, 802

\bibitem[{{de Zotti} et~al.(2005){de Zotti}, {Ricci}, {Mesa}, {Silva},
  {Mazzotta}, {Toffolatti}, \& {Gonz{\'a}lez-Nuevo}}]{dezotti/etal:2005}
{de Zotti}, G., {Ricci}, R., {Mesa}, D., {Silva}, L., {Mazzotta}, P.,
  {Toffolatti}, L., \& {Gonz{\'a}lez-Nuevo}, J. 2005, \aap, 431, 893

\bibitem[{{Dunkley} et~al.(2008)}]{dunkley/etal:prep}
{Dunkley}, J., et~al. 2008, ArXiv e-prints, 803

\bibitem[{{Finkbeiner}(2003)}]{finkbeiner:2003}
{Finkbeiner}, D.~P. 2003, \apjs, 146, 407, accepted (astro-ph/0301558)

\bibitem[{Finkbeiner et~al.(1999)Finkbeiner, Davis, \&
  Schlegel}]{finkbeiner/davis/schlegel:1999}
Finkbeiner, D.~P., Davis, M., \& Schlegel, D.~J. 1999, \apj, 524, 867

\bibitem[{{Gold} et~al.(2008)}]{gold/etal:prep}
{Gold}, B. et~al. 2008, \apjs

\bibitem[{Gorski et~al.(2005)Gorski, Hivon, Banday, Wandelt, Hansen, Reinecke,
  \& Bartlemann}]{gorski/etal:2005}
Gorski, K.~M., Hivon, E., Banday, A.~J., Wandelt, B.~D., Hansen, F.~K.,
  Reinecke, M., \& Bartlemann, M. 2005, \apj, 622, 759

\bibitem[{{Hill} et~al.(2008)}]{hill/etal:prep}
{Hill}, R. et~al. 2008, \apjs

\bibitem[{{Hinshaw} et~al.(2003)}]{hinshaw/etal:2003}
{Hinshaw}, G., et~al. 2003, \apjs, 148, 135

\bibitem[{{Hinshaw} et~al.(2007)}]{hinshaw/etal:2007}
---. 2007, \apjs, 170, 288

\bibitem[{{Hinshaw} et~al.(2008)}]{hinshaw/etal:prep}
---. 2008, ArXiv e-prints, 803

\bibitem[{{Huffenberger} et~al.(2006){Huffenberger}, {Eriksen}, \&
  {Hansen}}]{huffenberger/eriksen/hansen:2006}
{Huffenberger}, K.~M., {Eriksen}, H.~K., \& {Hansen}, F.~K. 2006, \apjl, 651,
  L81

\bibitem[{{Huffenberger} et~al.(2007){Huffenberger}, {Eriksen}, {Hansen},
  {Banday}, \& {Gorski}}]{huffenberger/etal:2007prep}
{Huffenberger}, K.~M., {Eriksen}, H.~K., {Hansen}, F.~K., {Banday}, A.~J., \&
  {Gorski}, K.~M. 2007, ArXiv e-prints, 710

\bibitem[{{Jarosik} et~al.(2007)}]{jarosik/etal:2007}
{Jarosik}, N., et~al. 2007, \apjs, 170, 263

\bibitem[{{Jones} et~al.(2006)}]{jones/etal:2006}
{Jones}, W.~C., et~al. 2006, \apj, 647, 823

\bibitem[{{Kamionkowski} et~al.(1997){Kamionkowski}, {Kosowsky}, \&
  {Stebbins}}]{kamionkowski/kosowsky/stebbins:1997}
{Kamionkowski}, M., {Kosowsky}, A., \& {Stebbins}, A. 1997, \prd, 55, 7368

\bibitem[{{Kogut} et~al.(2003)}]{kogut/etal:2003}
{Kogut}, A., et~al. 2003, \apjs, 148, 161

\bibitem[{{Komatsu} et~al.(2008)}]{komatsu/etal:prep}
{Komatsu}, E., et~al. 2008, ArXiv e-prints, 803

\bibitem[{{Leitch} et~al.(2005){Leitch}, {Kovac}, {Halverson}, {Carlstrom},
  {Pryke}, \& {Smith}}]{leitch/etal:2005}
{Leitch}, E.~M., {Kovac}, J.~M., {Halverson}, N.~W., {Carlstrom}, J.~E.,
  {Pryke}, C., \& {Smith}, M.~W.~E. 2005, \apj, 624, 10

\bibitem[{{Montroy} et~al.(2006)}]{montroy/etal:2006}
{Montroy}, T.~E., et~al. 2006, \apj, 647, 813

\bibitem[{{P.~Ade} et~al.(2007)}]{ade/etal:2007}
{P.~Ade}, et~al. 2007, ArXiv e-prints, 705

\bibitem[{{Page} et~al.(2007)}]{page/etal:2007}
{Page}, L., et~al. 2007, \apjs, 170, 335

\bibitem[{{Readhead} et~al.(2004)}]{readhead/etal:2004}
{Readhead}, A.~C.~S., et~al. 2004, \apj, 609, 498

\bibitem[{{Reichardt} et~al.(2008)}]{reichardt/etal:prep}
{Reichardt}, C.~L., et~al. 2008, ArXiv e-prints, 801

\bibitem[{Seljak \& Zaldarriaga(1997)}]{seljak/zaldarriaga:1997}
Seljak, U. \& Zaldarriaga, M. 1997, Phys. Rev. Lett., 78, 2054

\bibitem[{{Sievers} et~al.(2007)}]{sievers/etal:2007}
{Sievers}, J.~L., et~al. 2007, \apj, 660, 976

\bibitem[{Spergel \&
  Zaldarriaga(1997{\natexlab{a}})}]{zaldarriaga/spergel:1997}
Spergel, D.~N. \& Zaldarriaga, M. 1997{\natexlab{a}}, Phys. Rev. Lett., 79,
  2180

\bibitem[{Spergel \&
  Zaldarriaga(1997{\natexlab{b}})}]{spergel/zaldarriaga:1997}
---. 1997{\natexlab{b}}, \prl, 79, 2180

\bibitem[{{Toffolatti} et~al.(1998){Toffolatti}, {Argueso Gomez}, {de Zotti},
  {Mazzei}, {Franceschini}, {Danese}, \& {Burigana}}]{toffolatti/etal:1998}
{Toffolatti}, L., {Argueso Gomez}, F., {de Zotti}, G., {Mazzei}, P.,
  {Franceschini}, A., {Danese}, L., \& {Burigana}, C. 1998, \mnras, 297, 117

\bibitem[{{Wright} et~al.(2008)}]{wright/etal:prep}
{Wright}, E.~L. et~al. 2008, \apjs

\end{thebibliography}


\clearpage

\begin{deluxetable}{cccc}
\tablewidth{0pt}
\tablecaption{Unresolved Point Source Contamination\label{tbl:Aps}}
\tablehead{\colhead{Bands}&\colhead{Mask}
&\colhead{$A_{ps}(\alpha=0)$ [$10^{-3}\muK^2{\rm sr}$]}
&\colhead{$A_{ps}(\alpha=-0.09)$ [$10^{-3}\muK^2{\rm sr}$]}
}
\startdata
QVW & KQ85 & $11.3 \pm 0.9$ & $11.2\pm0.9$ \\
 & KQ80 & $11.3 \pm 0.9$ & $11.2\pm0.9$ \\
 & KQ75 & $10.7 \pm 1.0$ & $10.6\pm1.0$ \\
VW & KQ85 & $6.9\pm 3.4$ & $7.2\pm3.5$ \\
 & KQ80 & $9.1\pm 3.6$ & $9.5\pm3.8$ \\
 & KQ75 & $10.5\pm3.9$ & $11.1\pm4.1$ \\
\enddata
\tablecomments{All results are for $\ell=300-800$.}
\end{deluxetable}


\begin{deluxetable}{cccc}
\tablewidth{0pt}
\tablecaption{Unresolved Point Source Contamination\label{tbl:ptsrc}}
\tablehead{
&\colhead{$S_c$ [Jy]}
&\colhead{$C^{ps}$ [$10^{-3}\muK^2{\rm sr}$]}
&\colhead{$b^{ps}$ [$10^{-5}\muK^3{\rm sr}^2$]}}
\startdata
WMAP5 (KQ75)& \nodata & $11.7\pm1.1$\tablenotemark{a} & $4.3\pm1.3$\tablenotemark{b} \\
\cite{toffolatti/etal:1998}$\times0.64$ & 0.6 & 12.1 & 6.8 \\
& 0.5 & 10.4 & 4.9 \\
\cite{dezotti/etal:2005} & 0.7 & 11.7 & 8.4 \\
& 0.5 & 8.3 & 4.3 \\
\enddata
\tablenotetext{a}{By equation~\refeqn{eqn:srcmodel},
$C^{ps}(Q) = A_{ps}r(Q)^2 = 1.089\times A_{ps}$, where $A_{ps}$ is
the QVW/KQ75 result from \reftbl{tbl:Aps}.}
\tablenotetext{b}{From \cite{komatsu/etal:prep}, using the Q-band map and
KQ75 mask.}
\tablecomments{All numbers are evaluated at 40.7 GHz (Q band).}
\end{deluxetable}

\begin{deluxetable}{ccc}
\tablewidth{0pt}
\tablecaption{Beam/source likelihood treatment effect on parameters\label{tbl:srcnssig8}}
\tablehead{\colhead{Likelihood treatment}
&\colhead{$n_s$} & \colhead{$\sigma_8$}}
\startdata
standard & $0.964\pm  0.014$ & $0.796 \pm 0.036$ \\
SRCMARG & $0.964\pm  0.014$ & $0.798\pm  0.036$ \\
SRCMARG $\times5$ & $0.965\pm  0.015$ & $0.803\pm  0.042$ \\
BEAMMARG & $0.964\pm  0.015$ & $0.799\pm  0.036$ \\
BEAMMARG $\times20$ & $0.958\pm  0.016$  & $ 0.796\pm 0.034$ \\
\enddata
\tablecomments{
One-dimensional marginalized values for $n_s$ and $\sigma_8$ for various treatments
of the unresolved point source and beam uncertainty in the
{\WMAP} likelihood code.
See \refsec{sec:appsrclike} for descriptions of SRCMARG and BEAMMARG.
Here ``$\times5$'' and ``$\times20$'' indicate the error has been increased
by a factor of 5 and 20, respectively.
}
\end{deluxetable}


\clearpage

\begin{figure}
\plotone{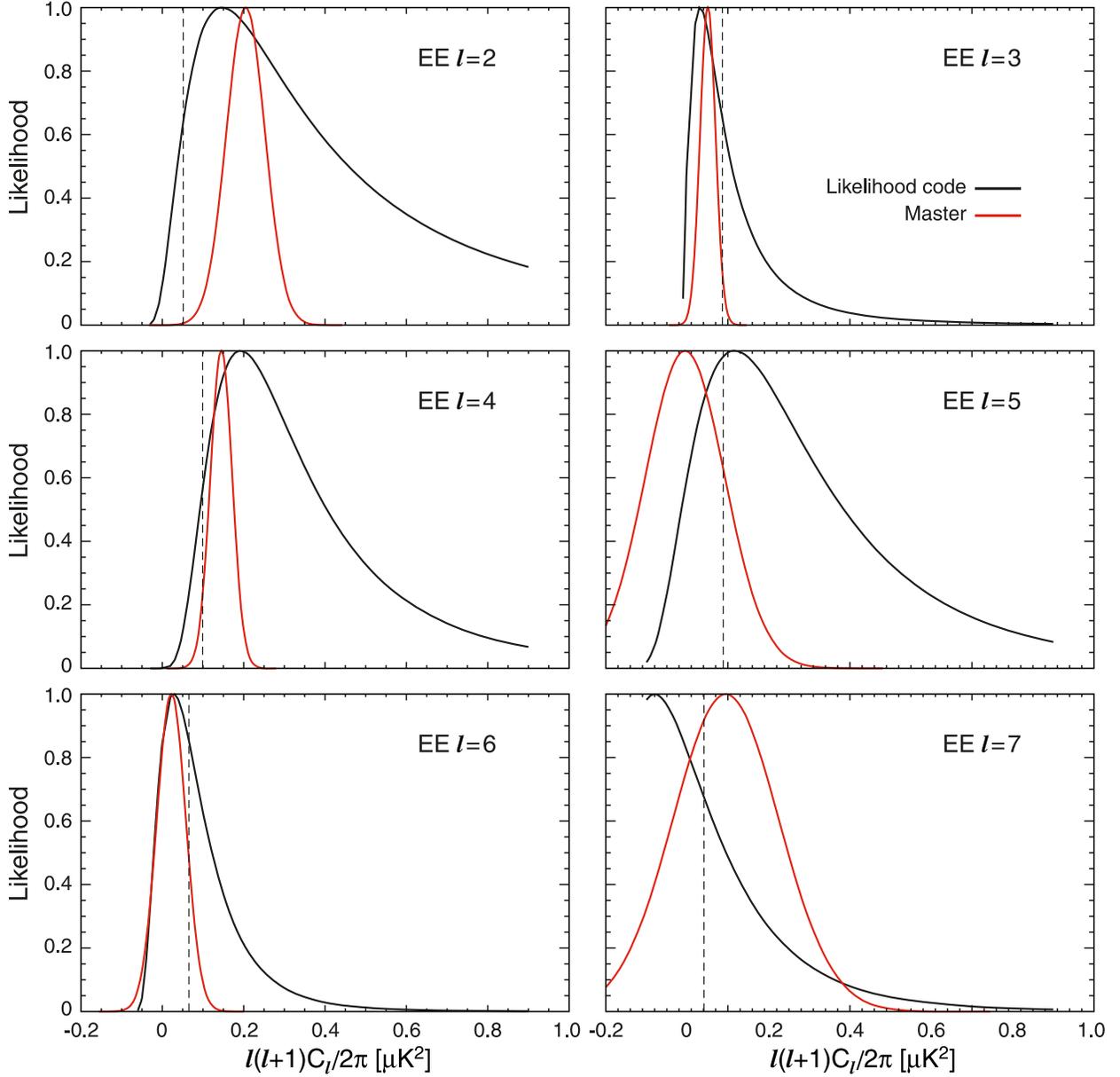} 
\caption{\label{fig:simpleee}
Conditional likelihoods for the $\ell=2-7$ EE multipole moments (black curves),
computed using the {\WMAP} likelihood code by varying the multipole in 
question, with all other multipoles fixed to their
fiducial values.
For example, in the $\ell=4$ panel, the black curve is
$f(x) \propto L(d|\dots,C^{EE}_3,C^{EE}_4=x,C^{EE}_5,\dots)$.
For comparison, na\"ive pseudo-$C_l$ estimates are also shown 
with Gaussian errors (red curves).
The pseudo-$C_\ell$ errors are noise only,
while the conditional distributions include cosmic variance.
}
\end{figure}

\begin{figure}
\plotone{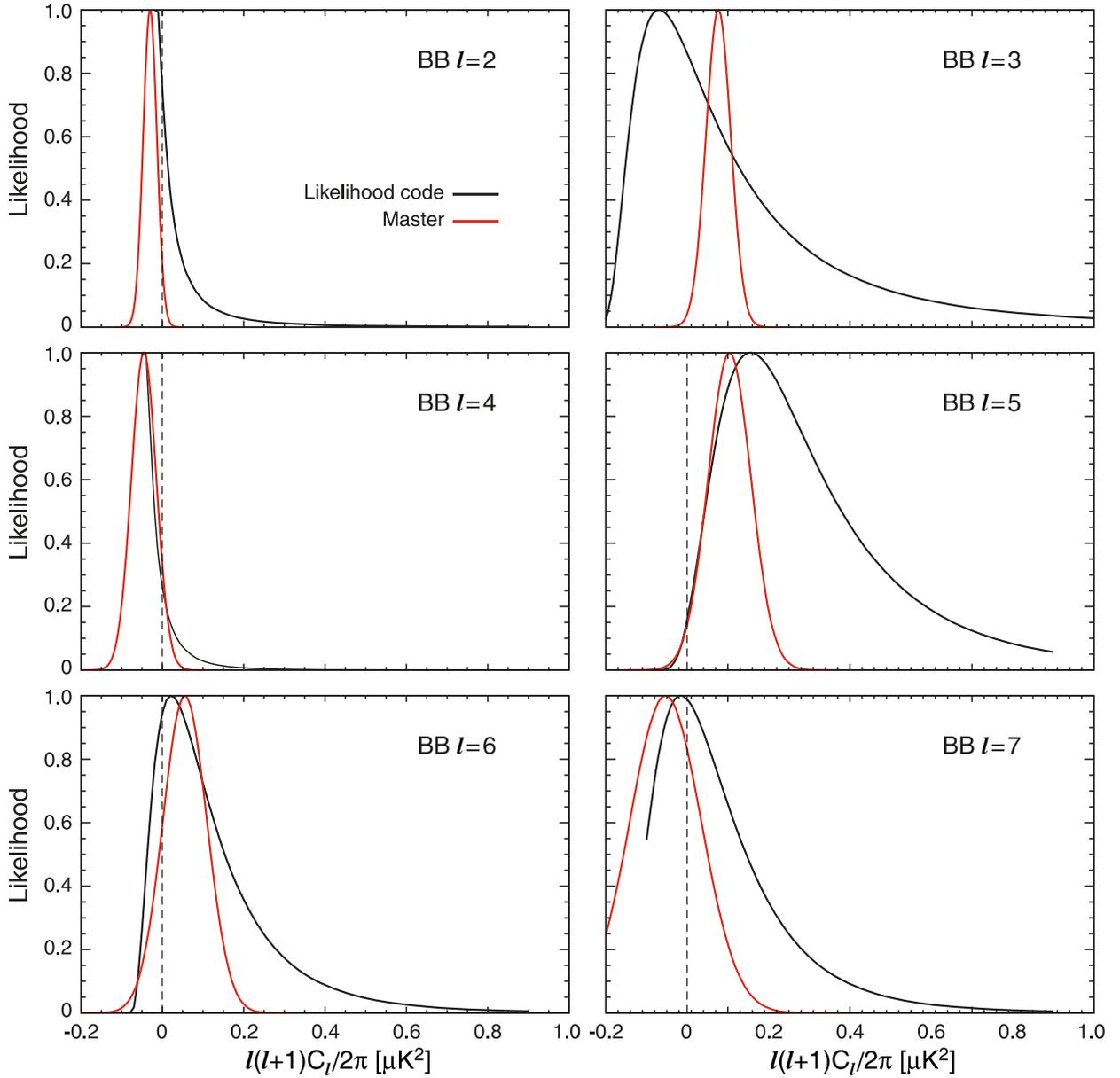} 
\caption{\label{fig:simplebb}
Conditional likelihoods for the $\ell=2-7$ BB multipole moments (black curves),
computed using the {\WMAP} likelihood code by varying the multipole in 
question, with all other multipoles fixed to their fiducial values.
For comparison, na\"ive pseudo-$C_l$ estimates are also shown
with Gaussian errors (red curves).
The pseudo-$C_\ell$ errors are noise only,
while the conditional distributions include cosmic variance.
Note the large difference between the likelihood code and 
the pseduo-$C_\ell$ value for $\ell=3$; this mode is sensitive to the time-orderd
data baseline and is extremely poorly measured by {\WMAP}, illustrating the 
complicated noise structure of the polarization data on large scales.
}
\end{figure}

\begin{figure}
\plotone{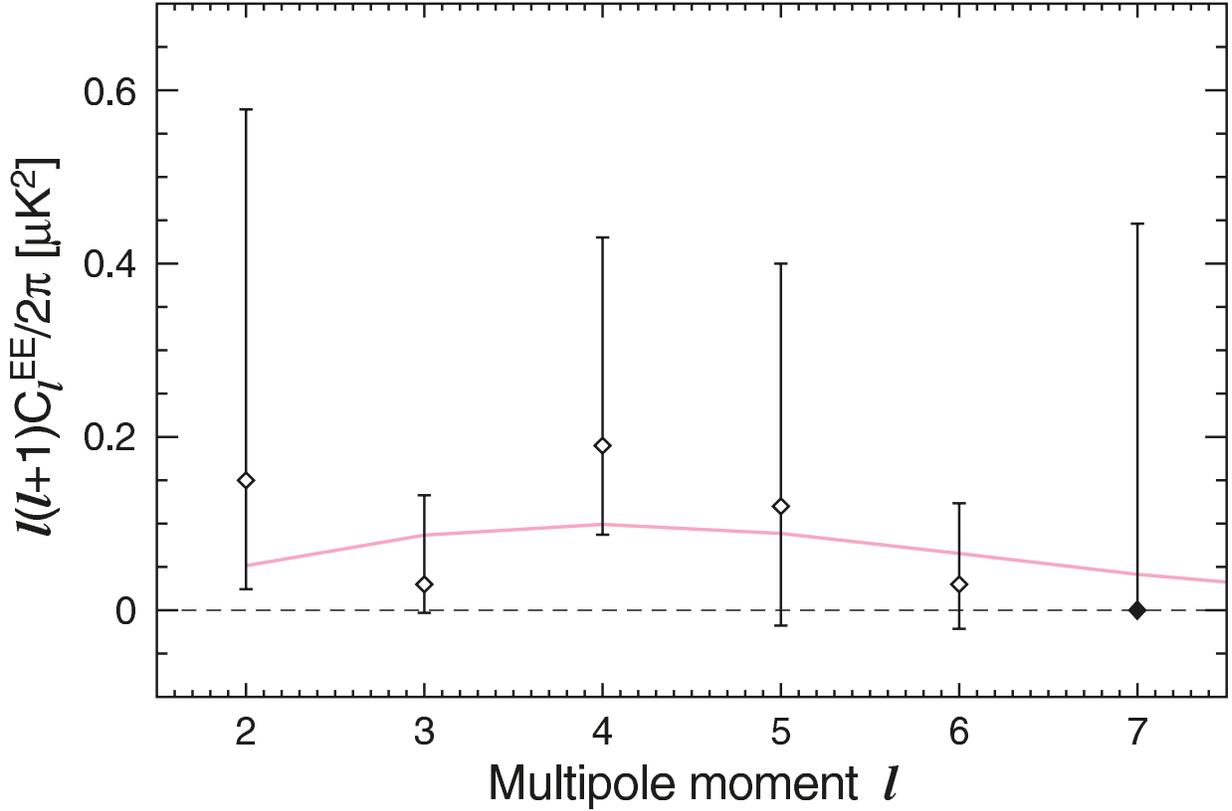}
\caption{\label{fig:eelowl}
{\WMAP} 5-year EE power spectrum at low-$\ell$.
The error bars are the 68\% CL of the conditional likelihood of
each multipole, with the other multipoles fixed at their fiducial theory values;
the diamonds mark the peak of the conditional likelihood distribution.
The error bars include noise and cosmic variance;
the point at $\ell=7$ is the 95\% CL upper limit.
The pink curve is the fiducial best-fit $\Lambda$CDM
model \citep{dunkley/etal:prep}.
}
\end{figure}

\begin{figure}
\plotone{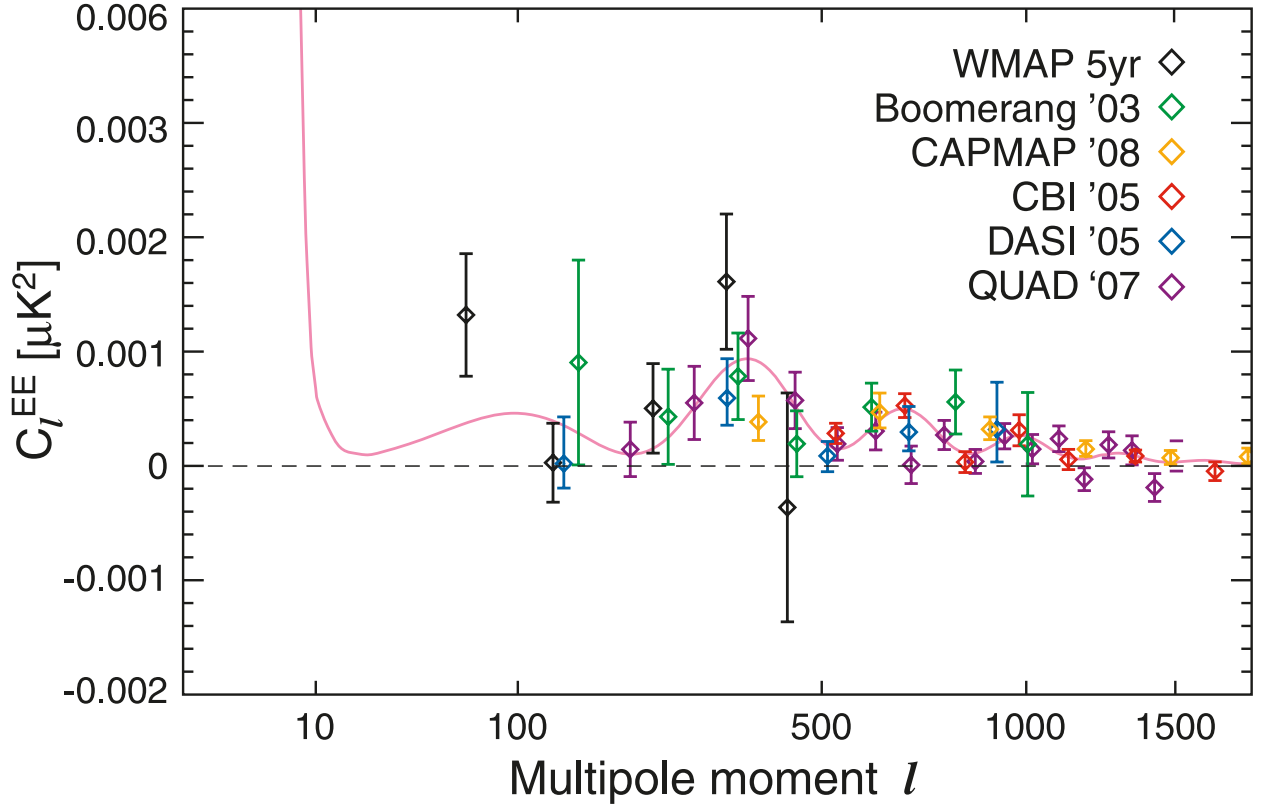}
\caption{\label{fig:eehighl}
{\WMAP} 5-year EE power spectrum, compared with results from the
Boomerang \citep[green]{montroy/etal:2006},
CAPMAP \citep[orange]{bischoff/etal:2008},
CBI \citep[red]{sievers/etal:2007},
DASI \citep[blue]{leitch/etal:2005},
and QUAD \citep[purple]{ade/etal:2007} experiments.
The pink curve is the best-fit theory spectrum from the $\Lambda$CDM/{\WMAP}
Markov chain \citep{dunkley/etal:prep}.
Note that the $y$-axis is $C^{EE}_\ell$,
not $\ell(\ell+1)C^{EE}_\ell/(2\pi)$.
}
\end{figure}

\begin{figure}
\plotone{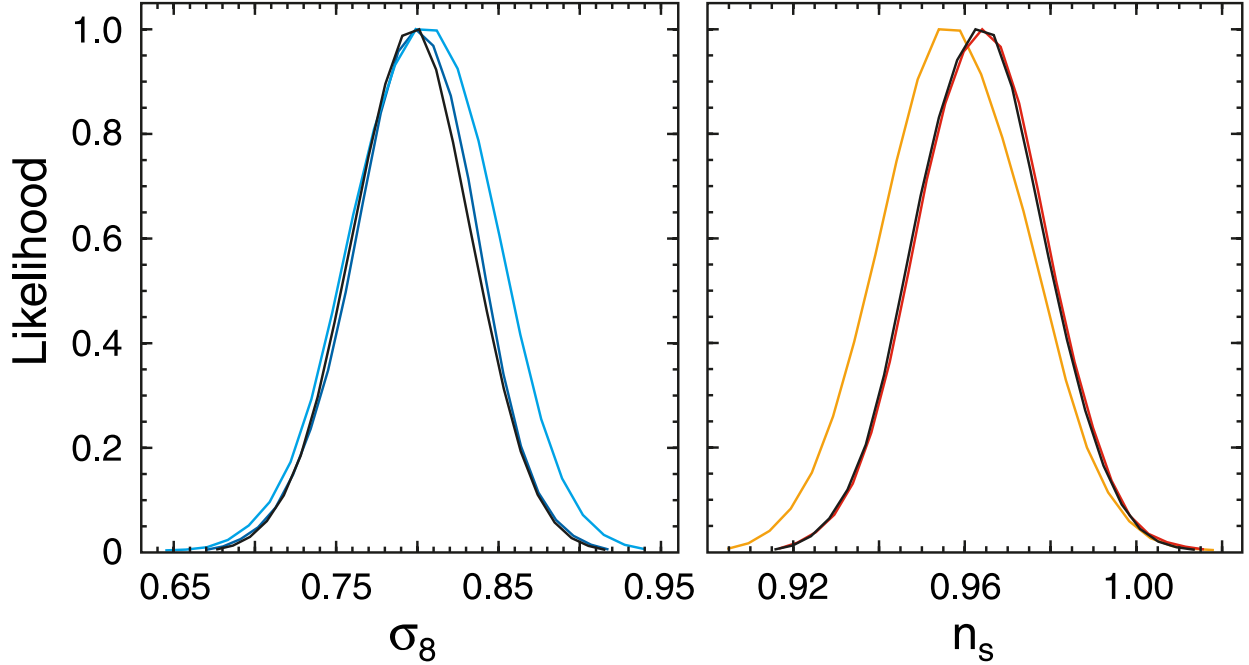}
\caption{\label{fig:nssig8}
\textit{Left:}
One-dimensional marginalized likelihood distributions of $\sigma_8$ for various treatments
of the source uncertainty in the likelihood code:
the standard likelihood function [black],
the alternative treatment of the source uncertainty described in
equation~\refeqn{eqn:srcquadrature} [blue],
the alternative treatment, but with the unresolved point source error
increased by
$\times5$ [cyan].
The agreement between black and blue curves shows that the standard treatment
is producing the correct answer.
\textit{Right:}
1D marginalized likelihood distributions of $n_s$ for various treatments
of the beam uncertainties:
the standard likelihood function [black],
the alternative treatment of the beam uncertainty described in
equation~\refeqn{eqn:srcquadrature} [red],
the alternative treatment, but with the beam error increased by
a factor of 20 [orange].
The agreement between the black and red curves shows that the standard
treatment is producing the correct answer.
}
\end{figure}

\begin{figure}
\plotone{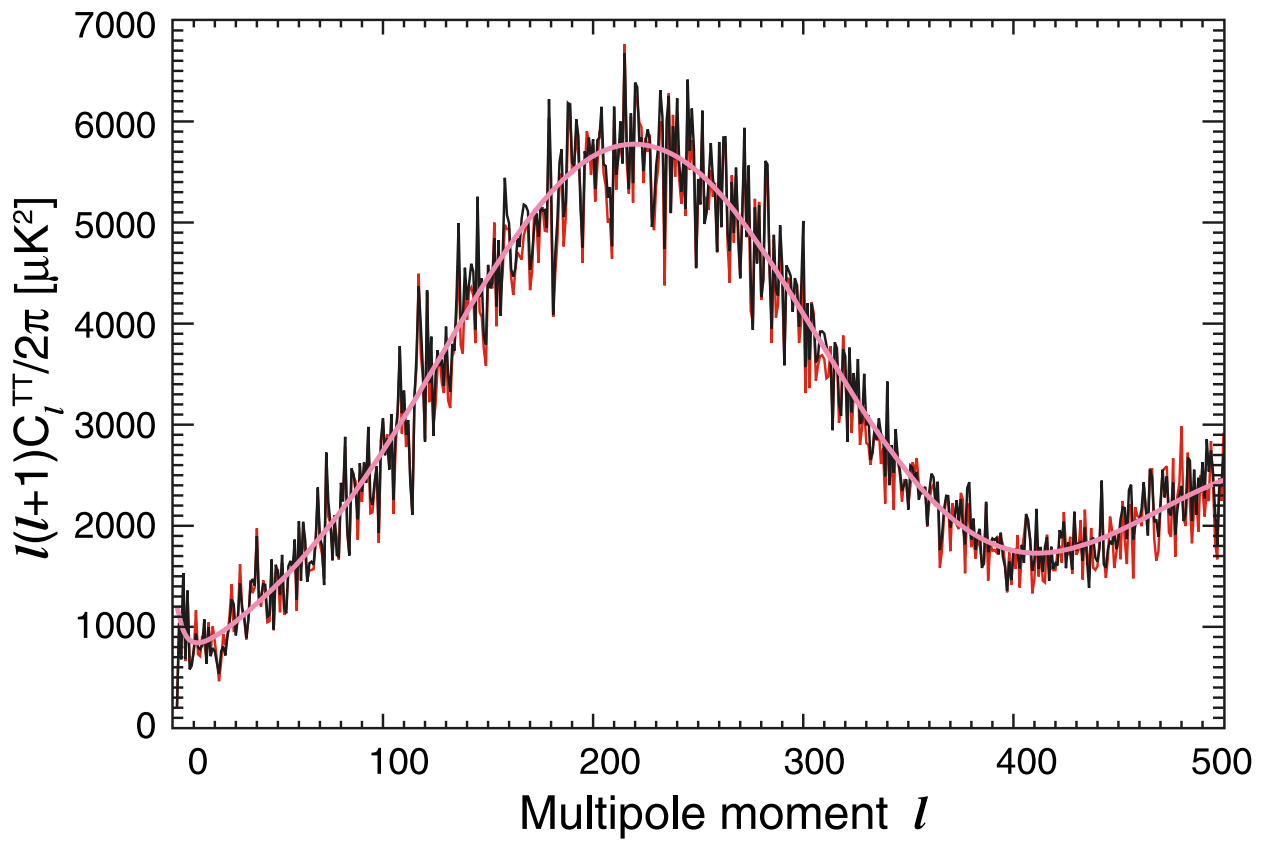}
\caption{\label{fig:ttunbinned}
The unbinned {\WMAP} 5-year temperature (TT) power spectrum (black),
compared with the {\WMAP} 3-year result (red). The slight upward shift 
of the 5-year spectrum relative to the 3-year spectrum is due to 
the change in the beam transfer function.
The pink curve is the best-fit $\Lambda$CDM model to the {\WMAP}5 data.
}
\end{figure}

\begin{figure}
\plotone{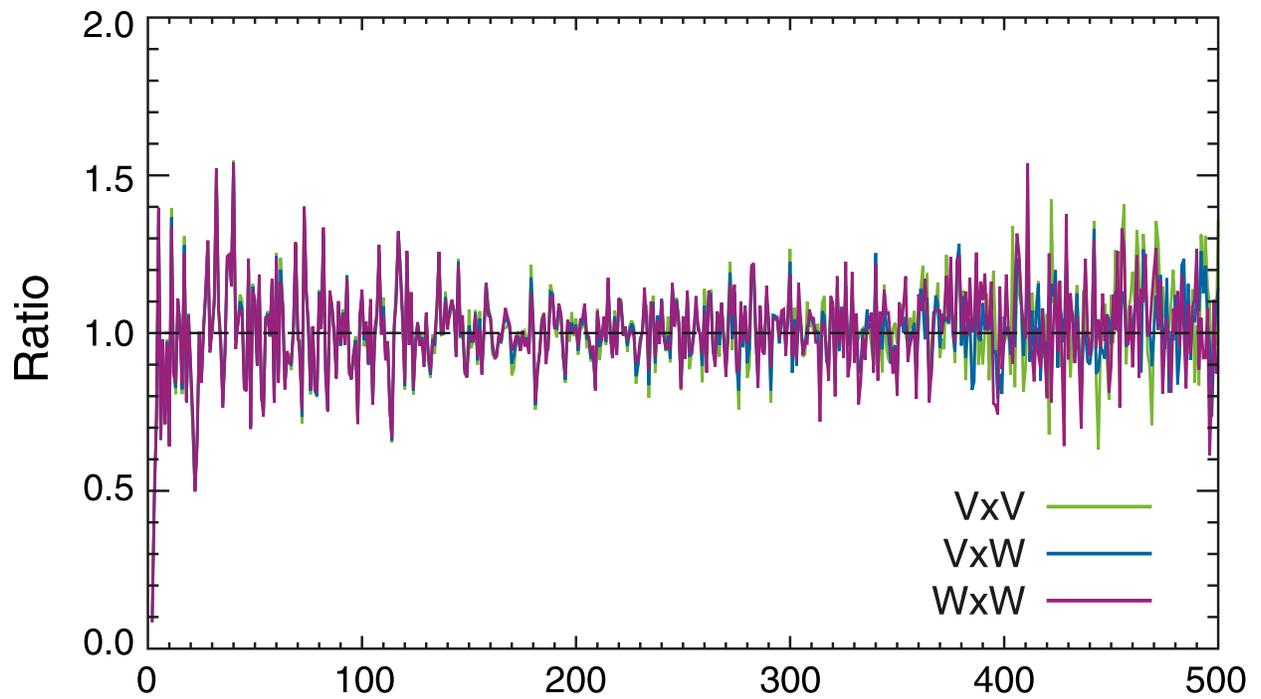}
\caption{\label{fig:ttfreq}
The unbinned {\WMAP} 5-year temperature (TT) power spectrum as a function
of frequency, divided by the best-fit $\Lambda$CDM model to the {\WMAP} data.
}
\end{figure}

\end{document}